# Large Fermi energy modulation in graphene transistors with high-pressure O$_2$-annealed Y$_2$O$_3$ topgate insulators

Kaoru Kanayama, *Kosuke Nagashio, Tomonori Nishimura, and Akira Toriumi

Department of Materials Engineering, The University of Tokyo 7-3-1 Hongo, Bunkyo-ku, Tokyo, 113-8656, Japan

*Email: nagashio@material.t.u-tokyo.ac.jp

We demonstrate a considerable suppression of the low-field leakage through a Y$_2$O$_3$ topgate insulator on graphene by applying high-pressure O$_2$ at 100 atm during post-deposition annealing (HP-PDA). Consequently, the quantum capacitance measurement for the monolayer graphene reveals the largest Fermi energy modulation ($E_F$ = ~0.52 eV, i.e., the carrier density of ~2×10$^{13}$ cm$^{-2}$) in the solid-state topgate insulators reported so far. HP-PDA is the robust method to improve the electrical quality of high-$k$ insulators on graphene.

The deposition of ultrathin and reliable high-$k$ dielectrics on graphene is required to realize the high transconductance in graphene field-effect transistors (FETs). Generally, two kinds of deposition methods are used: one is physical vapor deposition[1-4] (PVD) with low particle energies to avoid the defect introduction in graphene,[5] the other is atomic layer deposition (ALD) with buffer layers[6-8] to overcome the chemically inert surface of graphene. Recently, graphene FETs have been reported with ALD Al$_2$O$_3$ topgate insulators as thin as 2.6 nm.[9]

Insulators fabricated by PVD and ALD, however, suffer from the dielectric breakdown at the voltage much lower than the complete breakdown voltage due to the fragility of dielectrics accumulated by the low-field leakage during the iterative measurements. Typical electrical field for this low-field dielectric breakdown is ~0.2 V/nm. This is a critical issue for the reliability of the topgate insulators. A common limitation for the insulators on graphene fabricated by both PVD and ALD is the lack of a robust methodology for post-deposition annealing (PDA) in an O$_2$ atmosphere. Although PDA at a high temperature (e.g., ~500 °C) is known to improve the electrical quality of the insulator considerably,[10] it introduces many defects in graphene by oxidation.[11]

Here, from a thermodynamic viewpoint,[12] let's consider the Gibbs free energy change ($\Delta G = \Delta G° -$ RTln$P_{O2}$) for the oxidation reaction of M + O$_2$ = MO$_2$, where $\Delta G°$ is the standard Gibbs free energy change for the oxidation reaction, R is gas constant, $P_{O2}$ is the oxygen partial pressure and M is the metal. $\Delta G$ should be negative to facilitate the oxidation. The first term for $\Delta G$ can be used for material selection. An appropriate rare-earth element for use in high-$k$ insulators should be highly susceptible to oxidation. **Figure 1(a)** shows $\Delta G°$ of rare-earth, transition and representative elements at 300 °C calculated using a thermodynamic database.[13] The $\Delta G°$ for yttrium is negatively largest among all of the metal oxides, even smaller than that of carbon. Therefore, Y$_2$O$_3$ can be obtained at relatively low oxidation temperatures and is thermodynamically stable on graphene. The band gap for Y$_2$O$_3$ is ~5.5 eV, which is almost identical to that of h-BN.[14] The dielectric constant of Y$_2$O$_3$ is ~12,[15] while it is ~3-4 for h-BN.[16] The second term for $\Delta G$ is the process condition, which is derived from the oxygen potential. $\Delta G$ should be further decreased to reduce the defects in the insulator such as oxygen vacancy. This can be achieved by increasing $P_{O2}$ during PDA without elevating the annealing temperature.

The strategy for fabricating high-quality high-$k$ insulators on graphene is the deposition of Y$_2$O$_3$ on graphene by PVD with low particle energy and a subsequent high-pressure PDA (HP-PDA) process carried out in a 100 % O$_2$ atmosphere at ~100 atm and 300 °C. In this study, we demonstrate drastic suppression of the low-field leakage current through a Y$_2$O$_3$ topgate using HP-PDA. The improved electrical quality of the insulators on the monolayer graphene is realized by the large modulation of Fermi energy ($E_F$) in the quantum capacitance ($C_Q$) measurements as well as the high transistor performance in the electrical transport measurements.

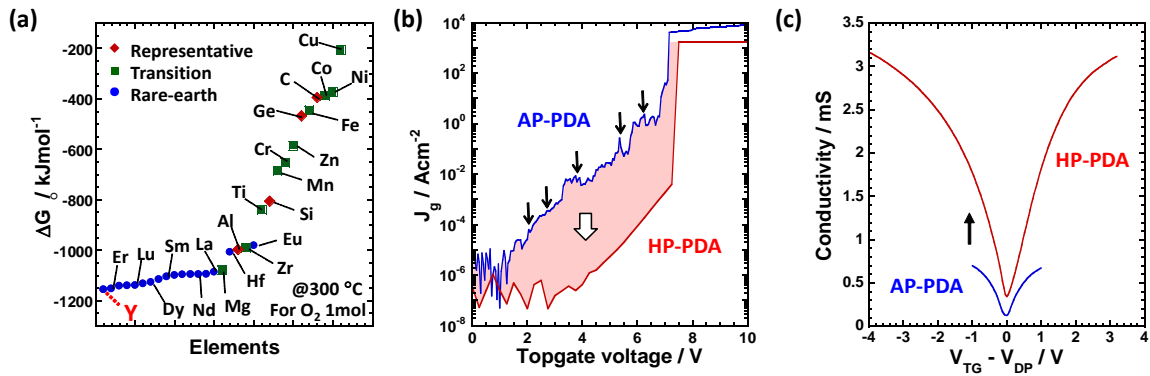

**FIG 1.** (a) $\Delta G°$ of rare-earth, transition and representative elements at 300 °C, calculated using a thermodynamic database. The $\Delta G°$ value for 1 mol of O$_2$ is shown, for mainly two kinds of oxidation reactions, M + O$_2$ = MO$_2$ and 4/3M + O$_2$ = 2/3M$_2$O$_3$. For the rare-earth elements, only several important elements are shown due to the lack of space. (b) Leakage current density through the Y$_2$O$_3$ topgate insulators fabricated by HP-PDA and AP-PDA. (c) $\sigma$ as a function of $V_{TG} - V_{DP}$ for the graphene FETs fabricated by HP-PDA and AP-PDA.

We fabricated monolayer graphene backgated FETs containing source and drain electrodes (Ni(~10 nm)/Au(~50 nm)) on ~90 nm $SiO_2$/n$^+$-Si substrates by the mechanical exfoliation of Kish graphite. $Y_2O_3$ was then deposited over the entire surface area of the wafer by the thermal evaporation of ~2 - 2.5 mg of Y metal in the $O_2$ atmosphere, where the $P_{O2}$ was adjusted to $10^{-1}$ Pa.[4] The thickness was roughly controlled at ~5 - 7 nm by selecting the initial mass of the Y metal. Next, the samples were placed in a high-pressure annealing system made of Inconel 625 alloy (supplementary material[17]), which is generally used for supercritical fluid experiments. The HP-PDA process was carried out in a 100 % $O_2$ atmosphere at ~100 atm at 300 °C for 20 min. Finally, the topgate electrode (Ni/Au) was patterned, followed by annealing at 300 °C for 30 s under 0.1% $O_2$ gas flow. For comparison with the HP-PDA process, atmospheric-pressure PDA (AP-PDA) was also carried out in a 100 % $O_2$ atmosphere at 1 atm at 200 °C for 10 min, which is the same condition used in the previous work.[4] The lack of a Raman D band measured through $Y_2O_3$ indicated that no defects were introduced into graphene by the HP-PDA process.[17]

**Figure 1(b)** compares the leakage current density ($J_g$) through $Y_2O_3$ between the source and the topgate fabricated by HP-PDA and AP-PDA. There are many small spikes in $J_g$ for AP-PDA, as shown by black arrows, suggesting that defect states exist within the band gap. After HP-PDA, the drastic suppression of the low field leakage current is clearly observed. Especially, small spikes completely disappear. On the other hand, the complete breakdown voltage does not change much. **Figure 1(c)** compares the conductivity ($\sigma$) as a function of the topgate voltage ($V_{TG}$) on the basis of the Dirac point voltage ($V_{DP}$) for the graphene FETs fabricated by HP-PDA and AP-PDA. It is evident that HP-PDA is effective for the transconductance improvement.

**Figure 2(a)** shows the resistivity ($\rho$) as a function of $V_{TG}$ for different backgate voltages ($V_{BG}$) at 20 K. The topgate leakage current of ~$10^{-12}$ A, as shown in the inset of **Fig. 2(a)**, is seven orders of magnitude lower than the source-drain current ($I_{SD}$) of ~$10^{-5}$ A. The topgate voltage can be swept over a wide range of ±4 V; therefore, the $V_{BG}$ is widely swept by ±60 V after the leakage current and complete breakdown voltage for the backgate $SiO_2$ insulator with ~90 nm in thickness are confirmed.

To qualitatively compare the electrical quality of the $Y_2O_3$ insulator with previous reports, we judged the insulator not from the relationship between $J_g$ and the equivalent oxide thickness (EOT) but rather from the maximum $E_F$ modulation estimated by the $C_Q$ measurement. **Figure 3(a)** shows the total capacitance ($C_{total}$) values between the source and topgate electrodes as a function of $V_{TG}$ for varying $V_{BG}$ at 20 K. The $C_{total}$ was strongly dependent on the $V_{TG}$. For channel materials with low density of states (*DOS*), such as graphene, the $C_{total}$ can be described by $1/C_{total} = 1/C_{ox} + 1/C_Q$, where $C_{ox}$ is the geometric capacitance and $C_Q = e^2 DOS$.[18] This is because the increase in the energy to induce carriers in the channels with low *DOS* is large and is introduced as a voltage drop in the equivalent circuit. The simplified equivalent circuit, in which $V_{DP}$ is held at $V_{TG} = 0$ V by adjusting the $V_{BG}$, is shown in **Fig. 3(b)**. The hysteresis in bidirectional C-V measurements is negligible for ±4 V sweep of $V_{TG}$,[19] as shown in **Fig. 3(c)**, suggesting the good quality of the $Y_2O_3$ insulator. **Figure 3(d)** compares the $C_{Total}$ as a function of $V_{TG}$ - $V_{DP}$ for the graphene FETs fabricated by HP-PDA and AP-PDA.[4] A considerable increase in the $C_{Total}$ was clearly observed, indicating an increase in $C_{Y2O3}$ by HP-PDA. This is consistent with the increase in the transconductance, as shown in **Fig. 1(c)**.

For qualitative estimation of the $C_{Y2O3}$, the $V_{DP}$ for the topgate sweep was plotted as a function of the $V_{BG}$, as shown in **Fig. 2(b)**. The DP shift of the I-V data is almost consistent with the shift of the C-V data. The average slopes of the I-V and C-V data for HP-PDA, which correspond to the capacitive coupling of $C_{SiO2}/C_{Y2O3}$, are more gradual than those observed for AP-PDA. Using $C_{SiO2} = 0.038$ μFcm$^{-2}$ for a 90 nm thickness of $SiO_2$, the $C_{Y2O3}$ for HP-PDA is calculated to be 0.95 µFcm$^{-2}$, which is indeed improved from $C_{Y2O3}$ = 0.76 µFcm$^{-2}$ for AP-PDA.[4] In the equivalent circuit shown in **Fig. 3(b)**, there is only one fitting parameter for the parasitic capacitance ($C_{para}$), as the $C_Q$ can be calculated by the theoretical relation of $C_Q = 2e^2 E_F/\pi(v_F \hbar)^2$, where $v_F$ is the Fermi velocity (1×10$^8$ cm/s) and $\hbar$ is Planck's constant. The experimental data are fit by solid lines in **Fig. 3(d)** using $C_{para}$ = 0.10 µFcm$^{-2}$ for HP-PDA and $C_{para}$ = 0.19 µFcm$^{-2}$ for AP-PDA. The $C_Q$ is obtained simply from $C_Q = C_{Y2O3} C'_{total}/(C_{Y2O3} - C'_{total})$, where $C'_{total} = C_{total} - C_{para}$. **Figure 3(e)** shows the $C_Q$ as a function of $E_F$ in the graphene FETs fabricated by HP-PDA. The $C_Q$ for AP-PDA is shown for comparison. The charging energy required to induce carriers in the graphene is denoted by $E_F$, which is expressed as $E_F = eV_{ch}$. $V_{ch}$ can be expressed by the equation for a series combination of capacitors according to $V_{ch} = V'_{TG} - \int_0^{V'_{TG}} C'_{Total}/C_{Y2O3}\, dV'_{TG}$, where $V'_{TG}$ is defined as $V'_{TG} = V_{TG} - V_{DP}$. The deviation from the theoretical value near the DP is due to residual carriers that are externally induced by the charged impurities.[2,4,20] The residual carrier density ($n^*$), as shown by the arrow in **Fig. 3(e)**, is calculated as 3.6×10$^{11}$ cm$^{-2}$ using $E_F = \hbar v_F \sqrt{\pi n^*}$. The $n^*$ values for HP-PDA and AP-PDA are almost the same, although there are device-to-device variations. The estimated $C_Q$ value for HP-PDA is consistent with the theoretical dotted line for $E_F >$ ~0.15 eV, suggesting that HP-PDA has no influence on the band structure of graphene. **Table 1** compares the maximum $E_F$ values reported in the literature. The data obtained by the ion gating are also included. Although there are many reports on electron transport in topgated graphene FETs, the literature for $C_Q$ measurements is selected because relatively high-quality and thin topgate insulators are required. The largest $E_F$ values (0.52 eV) for solid-state topgate insulators are achieved by HP-PDA.

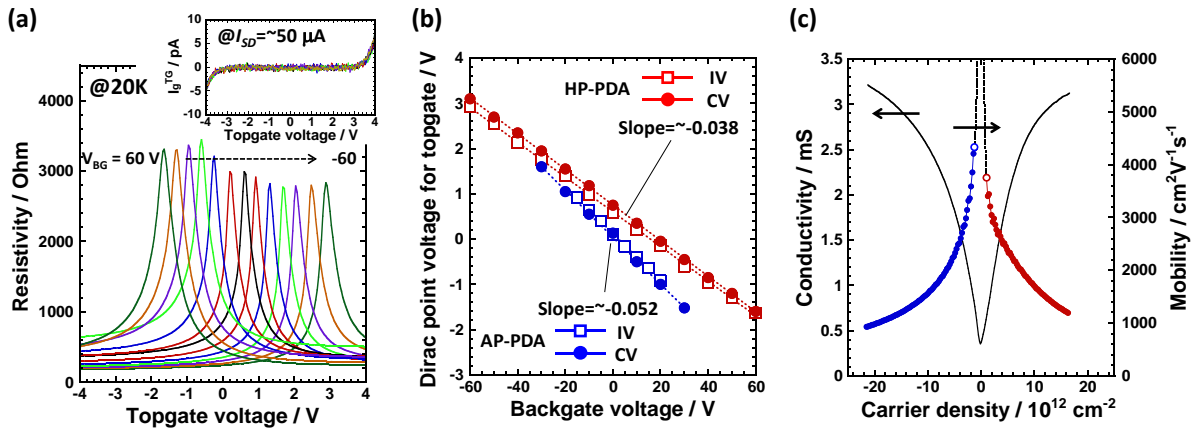

**FIG 2.** (a) $\rho$ as a function of $V_{TG}$ for different $V_{BG}$ at 20 K. The inset shows the topgate leakage current for $I_d$ = ~50 μA. (b) $V_{DP}$ for the $V_{TG}$ sweep as a function of the $V_{BG}$. The red solid circles and red solid boxes indicate the DPs obtained by I-V and C-V measurements for HP-PDA in Fig. 2(a) and Fig. 3(a), respectively. The blue data are adapted from a previous work.[4] The slopes are determined by linear fitting. (c) $\sigma$ and $\mu$ as a function of $n$. The mobility for $n < 1\times10^{12}$ cm$^{-2}$ is not valid due to $n^*$.

Now, let's move back to the transport measurement to extract the carrier mobility ($\mu$). The carrier density ($n$) cannot be evaluated by a simple parallel plate capacitor model for geometric capacitance, because of the significant contribution of the $C_Q$. Here, $n$ is evaluated by the integral of the differential capacitance $C'_{total}$,[21] i.e., $n = \frac{1}{e} \int_0^{V'_{TG}} C'_{Total}\, dV'_{TG}$. Then, $\mu$ is calculated from $\mu = \sigma/en$. **Figure 2(c)** shows $\sigma$ and $\mu$ as a function of $n$. A large $n$ of ~2×10$^{13}$ cm$^{-2}$ is reached because of the improved $C_{Y2O3}$ and the large voltage applied only to the $Y_2O_3$ insulator ($V_{TG} - V_{ch}$ = ~3.6 V). Although $n$ is evaluated by the integral of $C'_{total}$, it still includes the contribution from $n^*$ near the DP. Therefore, $\mu$ for $n < 1\times10^{12}$ cm$^{-2}$ is neglected. The high $\mu$ of ~4,100 cm$^2$V$^{-1}$s$^{-1}$ is much improved from the previous AP-PDA device,[4] even though the contribution of the contact resistance is included in this device.

Next, we discuss the physical and chemical origins for the drastic improvement of the $Y_2O_3$ insulator fabricated by HP-PDA. The X-ray photoelectron spectra were measured for the $Y_2O_3$ films on highly oriented pyrolytic graphite (HOPG) with HP-PDA and AP-PDA.[17] The peaks for $C_{1s}$, $Y_{3d}$ and $O_{1s}$ are almost the same for HP-PDA and AP-PDA. Therefore, the oxidation state of the Y element is not drastically affected by HP-PDA. Moreover, the average surface roughness of the $Y_2O_3$ film on the HOPG with HP-PDA and AP-PDA are ~0.36 nm and ~0.34 nm, respectively, again suggesting no clear difference.

The main difference is evident in the cross-sectional transmission electron microscope (TEM) images of the Y$_2$O$_3$ topgate graphene devices shown in **Fig. 4(a)**. The Y$_2$O$_3$ insulator fabricated by HP-PDA is crystallized as a cubic phase (the inset in **Fig. 4(a)**); however, an amorphous region is also detected around the crystallized grain. On the other hand, the Y$_2$O$_3$ insulator fabricated by AP-PDA has very limited partial crystallization.[22] The increase in the dielectric constant by HP-PDA is explained by the higher dielectric constant of crystalline Y$_2$O$_3$. Interestingly, the formation of a two-dimensional hexagonal Y$_2$O$_3$ monolayer on graphene has been recently reported.[23] However, we did not observe a specific crystallographic relationship between Y$_2$O$_3$ and graphene. Moreover, electron energy loss spectroscopy (EELS) detected the NiO$_x$ region on the Y$_2$O$_3$ insulator, which may have been formed by the annealing at 300 °C under the 0.1% O$_2$ gas flow after the topgate electrode deposition, as shown in **Fig. 4(b)**. This observation explains the apparently low $k$ value of 5.4 for the HP-PDA Y$_2$O$_3$ insulator with a C$_{Y2O3}$ = 0.95 mFcm$^{-2}$ and a thickness of 5 nm. The topgate insulator in this study should be recognized as a gate stack of NiO$_x$ and Y$_2$O$_3$ double layer.

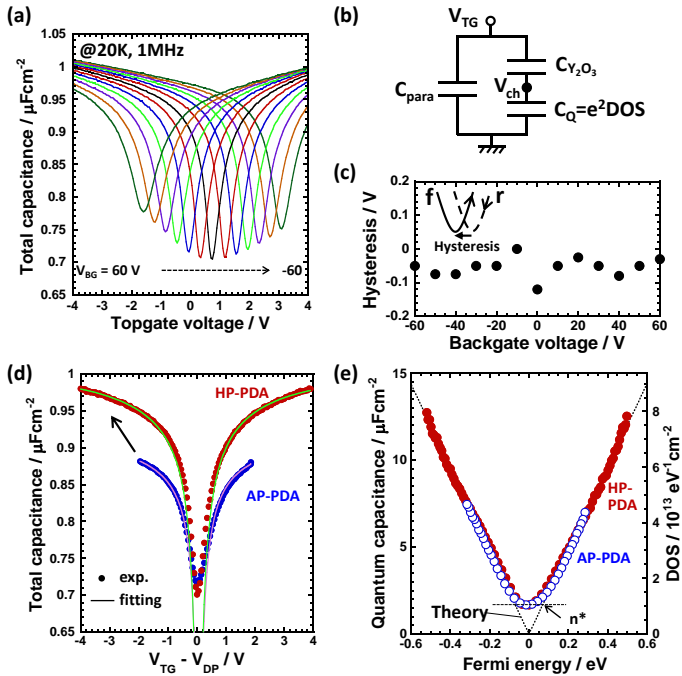

**FIG 3.** (a) $C_{Total}$ as a function of $V_{TG}$ for the graphene FET device at 20 K at a frequency of 1 MHz. (b) The equivalent circuit simplified for the topgate graphene FET device, in which $V_{DP}$ is placed at $V_{TG} = 0$ V by adjusting the $V_{BG}$. (c) Hysteresis detected in the sweeping range of $\Delta V_{TG} = \pm 4$ V for different $V_{BG}$. The hysteresis is defined as $\Delta V_{hys} = V_{DP}^{forward} - V_{DP}^{reverse}$ for the $V_{TG}$ sweep. (d) Comparison of $C_{Total}$ as a function of $V_{TG} - V_{DP}$ for the graphene FET devices fabricated by HP-PDA and AP-PDA.[4] Solid circles are experimental data, while solid lines are curves fitted by $C_{para}$ = 0.10 µFcm$^{-2}$ for HP-PDA and $C_{para}$ = 0.19 µFcm$^{-2}$ for AP-PDA. (e) $C_Q$ extracted based on the fitting in (d). The dotted line is the theoretical $C_Q$. The right vertical axis is converted to the *DOS*. $n^*$ is 3.6×10$^{11}$ cm$^{-2}$ for both HP-PDA and AP-PDA.

Finally, we discuss the role of graphene as a diffusion barrier during HP-PDA. **Figure 4(c)** shows the capacitance as a function of the gate voltage for Au/Y$_2$O$_3$/n-Si MOS capacitors fabricated by AP-PDA and HP-PDA. The considerable degradation of the capacitance for the accumulation side for HP-PDA indicates that the decrease in the $k$ value for the Y$_2$O$_3$ insulator is due to the diffusion of Si into the Y$_2$O$_3$, which has been reported previously.[24] On the other hand, the $C_{total}$ for Y$_2$O$_3$ on graphene increases for HP-PDA, as shown in **Fig. 3(d)**. Si diffusion in Y$_2$O$_3$ is not detected within the resolution of the EELS measurements shown in **Fig. 4(b)**. Although the tail of the composition profile is generally ~2 nm in length due to the spatial extent of the electron beam, the lack of Si at the center of the Y$_2$O$_3$ profile and similar slopes for both the Y and O profiles support our premises. These results suggest that the increase in $P_{O2}$ by HP-PDA are totally spent on the improvement of the crystallinity and the reduction of defects such as oxygen vacancies, rather than on the diffusion of the other elements, because graphene is known to work as a diffusion barrier,[25-27] as schematically explained in **Fig. 4(d)**. HP-PDA is applicable to other high-$k$ materials on graphene, because the difference in $\Delta G°$ is very small for high-$k$ materials. Moreover, this method may be extended to other layered materials with relatively high melting points.

HP-PDA is the robust method to drastically improve

**Table 1** Summary of maximum $E_F$ in the $C_Q$ measurements

| Insulator | Deposition method | Graphene | $E_F$ [eV] | $n$ [cm$^{-2}$] | Thickness [nm] | $C_{ox}$ [μFcm$^{-2}$] | Ref. |
|---|---|---|---|---|---|---|---|
| Al$_2$O$_3$ | ALD | ME* | 0.3 | ~6.5 × 10$^{12}$ | 10 | 0.56 | 28 |
| Al$_2$O$_3$ | Depo. in O$_2$ | ME | 0.35 | ~10 × 10$^{12}$ | 10 | 0.47 | 2 |
| Y$_2$O$_3$ | Metal depo. & Air anneal | ME | 0.4 | ~10 × 10$^{12}$ | 3.9 | 1.89 | 1 |
| HO$_2$ | ALD** | CVD | 0.3 | ~6.5 × 10$^{12}$ | 4.4 | 1.98 | 29 |
| h-BN | ME & Transfer | ME | 0.26*** | ~5 × 10$^{12}$ | 27 | 0.15 | 30 |
| Y$_2$O$_3$ | Depo. in O$_2$ & HP-PDA | ME | 0.52 | ~20 × 10$^{12}$ | - | 0.95 | present |
| BMIM-PF$_6$ | Ion gate | ME | 0.22 | ~3.5 × 10$^{12}$ | ~0.3 | ~21 | 31 |
| ABIM-TFSI | Ion gate | ME | 0.82*** | ~50 × 10$^{12}$ | | | 32 |
| Solid polymer | Ion gate | ME | 1.17*** | ~100 × 10$^{12}$ | | ~3.2 | 33 |

*"ME" indicates mechanical exfoliation of bulk graphite. **ALD HfO$_2$ is deposited on embedded metal and CVD graphene is transferred on HfO$_2$. The device has inverted structure. ***E$_F$ for these cases was calculated form the carrier density.

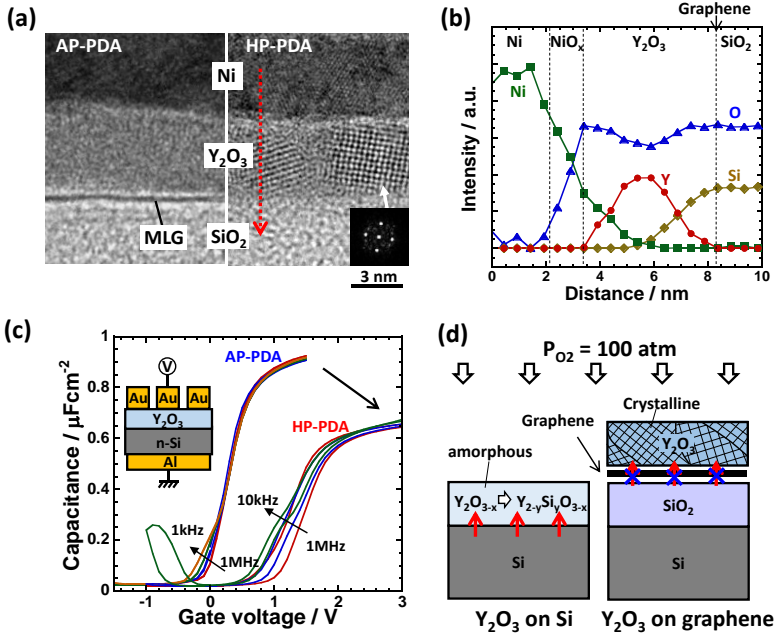

**FIG 4.** (a) Cross-sectional TEM images for the graphene FET devices fabricated by HP-PDA and AP-PDA. The inset shows the Fourier transform of the Y$_2$O$_3$ grain marked by an arrow, confirming the cubic structure. (b) Intensity profiles as a function of distance along the dotted red line in (a), determined by the EELS measurements. It should be noted that the full-width at half-maximum for the Y profile is narrower than the thickness of Y$_2$O$_3$ determined from the TEM image because the detection sensitivity for the Y element is lower than for the other elements. (c) Bi-directional $C$-$V$ characteristics for the Au/Y$_2$O$_3$/n-Si MOS capacitor fabricated by HP-PDA and AP-PDA at RT. (d) Schematics of the effect of HP-PDA for Y$_2$O$_3$ on Si (left) and Y$_2$O$_3$ on graphene/SiO$_2$/Si (right).

the electrical quality of high-*k* insulators without elevating the annealing temperature. The key to this process is that the increase in $P_{O2}$ by HP-PDA is totally spent on the improvement of crystallinity and the reduction of defects such as oxygen vacancies, rather than on the diffusion of other elements, because graphene works as a diffusion barrier. Using the high-quality $Y_2O_3$ insulator, we demonstrate the largest $E_F$ (0.52 eV) reported to date in solid-state topgate insulators as well as the high transistor performance.

We are grateful to Covalent Materials for kindly providing the Kish graphite. K. N. acknowledges financial support from the JSPS through its "Funding Program for World-Leading Innovative R&D on Science and Technology (FIRST Program)" and from a Grant-in-A id for Scientific Research on Innovative Areas and for Young scientists by the Ministry of Education, Culture, Sports, Science and Technology in Japan.